\documentstyle[prb,aps,epsfig,latexsym,floats,twocolumn]{revtex}

\newcommand{\bk}{{\bf k}}
\def\kf{k_{\text F}}
\newcommand{\br}{{\bf r}}
\newcommand{\mbf}[1]{{\mathbf #1}}
\newcommand{\ofr}{{(\mbf{r})}}       % (r vec)
\newcommand{\GT}{\Gamma_{\text{T}}}
\newcommand{\DEL}{\mbox{\boldmath $\nabla$}}
\newcommand{\LB}{Landauer-B\"{u}ttiker}
\def\lf{\lambda_{\text F}}
\def\st{\sigma_{\text T}}
\def\stlr{\sigma_{\text T}^{\text{L$\rightarrow$R}}}
\def\strl{\sigma_{\text T}^{\text{R$\rightarrow$L}}}
\def\aeff{a_{\text{eff}}}
\def\aaeff{A_{\text{eff}}}
\def\gat{G_{\text{atom}}}
\newcommand{\be}{\begin{equation}}
\newcommand{\ee}{\end{equation}}
\newcommand{\bea}{\begin{eqnarray}}
\newcommand{\eea}{\end{eqnarray}}
\def\bean{\begin{mathletters}\begin{eqnarray}}
\def\eean{\end{eqnarray}\end{mathletters}}
\newcommand{\etal}{{\it et al.\ }}

\def\eg{{\it e.g.\ }}
\def\ie{{\it i.e.\ }}
\newcommand{\half}{\mbox{\small $\frac{1}{2}$}}
\newcommand{\pit}{\mbox{\small $\frac{\pi}{2}$}}

\begin{document}

\draft

\title{Mesoscopic scattering in the half-plane:
squeezing conductance through a small hole}

\author{A. H. Barnett, M. Blaauboer, A. Mody and E. J. Heller}
\address{Department of Physics, Harvard University, Cambridge MA 02138
}
\date{\today}
\maketitle

\begin{abstract}

We
model the 2-probe  conductance of a quantum
point contact (QPC), in linear response.
%connecting two open reservoir regions 
%of two-dimensional electron gas.
%in the linear response regime.
%
If the QPC is highly non-adiabatic or near to
scatterers in the open reservoir regions, then
the usual distinction between leads and reservoirs breaks down and
a technique based on scattering theory in the full
two-dimensional half-plane
is more appropriate.
Therefore we
relate conductance to the transmission {\em cross section}
for incident plane waves.
This is equivalent to Landauer's formula using a radial partial-wave basis.
We derive the
result that an
arbitrarily small (tunneling) QPC can reach a p-wave channel conductance
of $2e^2/h$ when coupled to a suitable reflector.
If two or more resonances coincide the total conductance can even exceed this.
This relates to recent mesoscopic experiments in open geometries.
We also discuss reciprocity of conductance, and the possibility
of its breakdown in a proposed QPC for atom waves.

\end{abstract}


%%%%%%%%%%%%%%%%%%%%%%%%%%%%%%%%%%%%%%%%%%%%%%%%%%%%%%%%%%%%%%%%%%%%%%%%%%%%%%%%
\section{Introduction}

The quantum point contact\cite{been,dittrich} (QPC)
has played a
central role in the understanding of mesoscopic conductance.
It is the simplest example of a 2DEG system
where the quantum coherent nature of the electron controls the bulk
transport properties.
The \LB\ (LB) formalism
\cite{LB,datta,dittrich}
reduces the calculation of quantum conductance
in the linear response regime
to the evaluation of single-particle wavefunction transmission amplitudes.
Traditionally, these amplitudes are measured between
travelling wave basis states in the `leads'.
Far from the scattering system the leads
have constant profiles of
finite-width, and support a finite number of
transverse modes (channels).
Eventually it is assumed that the leads are impedance-matched
(that is, without
reflection) into `reservoirs'
which act as thermalized sources of electrons at their respective
potentials; these potentials are taken to reflect the measured
bias voltage.
Such theoretical constructs have been remarkably successful at describing
%2-terminal experimental
transport
phenomena,
for instance conductance quantization \cite{wees88,been,dittrich}, because
the scattering systems involved have generally had
good lead-to-reservoir matching.
% what about qdots?

We consider `open' 2-terminal mesoscopic systems,
namely those
where a QPC is {\em non-adiabatic} 
(possessing rapid longitudinal variation in
transverse profile \cite{nonadiab,been}) and has short or
nonexistent leads
(for instance if it suddenly abutts onto the `reservoir' regions),
or those where there can be scattering off nearby
objects in the `reservoir' region.
We call such systems `open' because the fully two-dimensional (2D)
nature of the `reservoirs'
(\ie the surrounding semi-infinite regions of free space) is important,
and therefore they cannot be modelled using the quasi-1D approach described
above.
This includes a variety of recent
mesoscopic experiments, for example the combination of QPCs with nearby
resonator structures \cite{kati97} or with a nearby
depletion region underneath
an AFM tip \cite{topinka}.
It also includes any QPC system where elastic backscattering from
disorder in the reservoirs is significant\cite{geim94}, or generally where the
lead-reservoir matching is bad.
In such systems, the conventional quasi-1D picture
does not apply: the scattering system is not coupled to leads
in the usual sense, indeed the
distinction between leads and reservoirs
is no longer clear \cite{szafer}.
The main aim of the present work is to
introduce a 2D scattering theory approach
which can handle such systems, and to apply it to the calculation
of the maximum conductance of an
open resonator structure of experimental
relevance.

\begin{figure}[t]
\centerline{\epsfig{figure=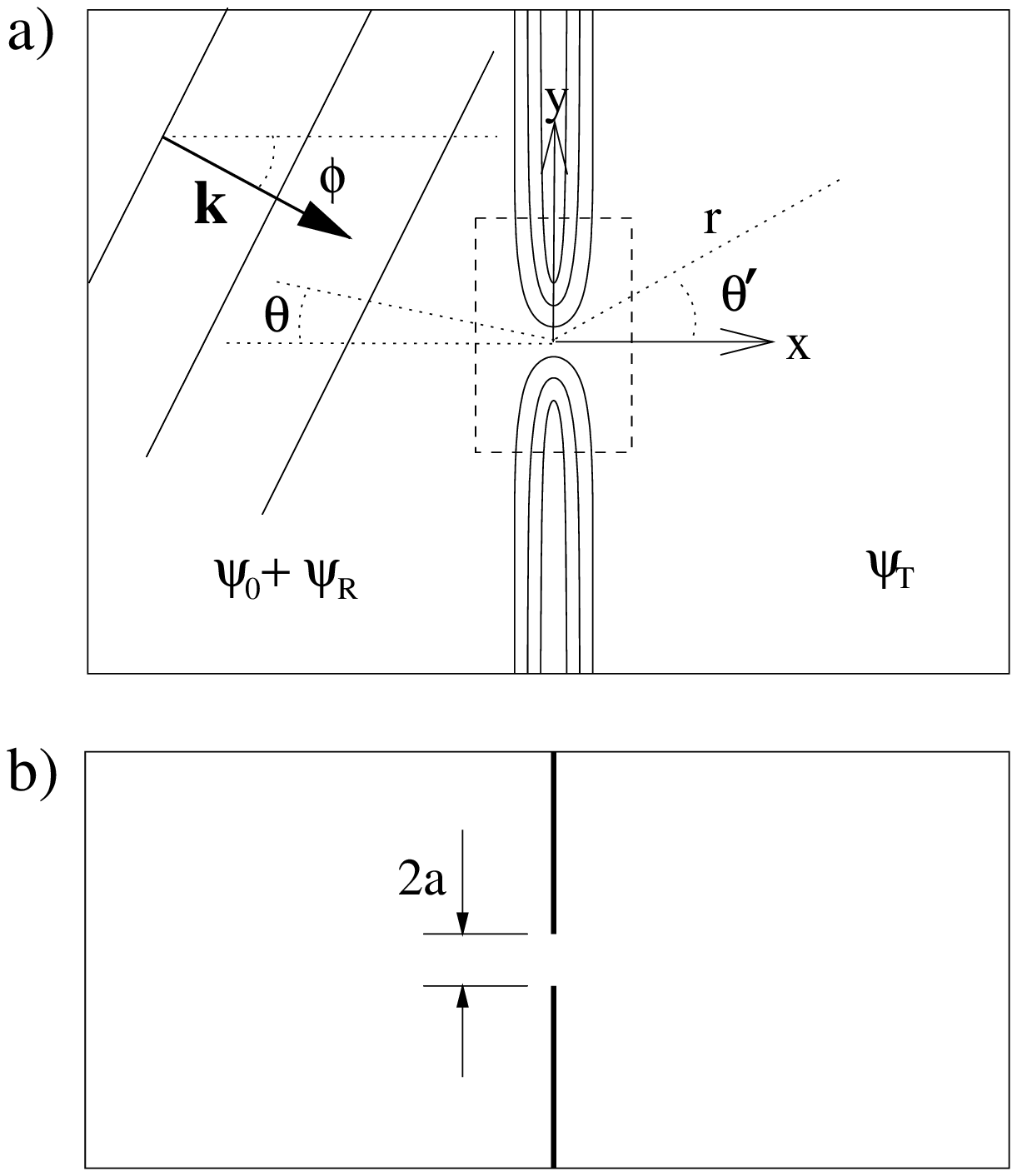,width=0.9\hsize}\vspace{.1in}}
\caption{
Schematic QPC geometry in 2D:
{\bf a)} general point 
contact scatterer coupling two semi-infinite regions of free space.
The solid curves are contours of an elastic scattering potential $V\ofr$.
The `system' size $L$ (dashed box) we take to be the region where $V\ofr$
has not yet reached its asympotic form
(which is zero apart from a $y$-invariant profile around the $y$-axis).
Also shown are
an incoming plane wave,
and the coordinate system.
{\bf b)} The idealized `slit' aperture in a thin, hard wall
considered in Section~\protect\ref{sec:coupled}.
}
\label{fig:geom}
\end{figure}

We imagine
a geometry where a 2DEG exists in two semi-infinite
half-plane
regions, separated by an impenetrable potential barrier which we
align with the $y$-axis
(see Fig. \ref{fig:geom}a).
Our general `QPC scattering system'
is any gap in this barrier which allows coupling of the
wavefunction on the left and right sides.
This gap can be defined by an arbitrary form of the elastic potential,
and may include other nearby scattering objects or disorder
(which would all be placed within the boxshown in Fig.~\ref{fig:geom}a).
The only important limitation is that this coupling region
(the `system') be of finite
$y$ extent, so that electrons 
which leave
the system do so via a well-defined terminal:
either the left ($x<0$) or 
the right ($x>0$).
We also assume that the system size $L$
is much smaller than both the dephasing length $l_\phi$ and
the momentum relaxation (elastic scattering) length $l_e$.
The former requirement allows treatment using a coherent wavefunction
across the system; the latter allows free-space elastic scattering concepts
to be applied.
We will stay within the non-interacting quasiparticle picture,
consider zero applied magnetic field, and assume spin degeneracy of 2
throughout.

The conventional distinction between `reservoir' and `lead'
is no longer applicable, however
at short distances outside the system
($r > L$ but $r \ll l_\phi$ and $r \ll l_e$) the two semi-infinite
free space regions behave like leads, since they support scattering-free
`channels' (see Section~\ref{sec:pw}).
At large distances the {\em same} regions behave as reservoirs:
for $r \gg l_e$ ergodicity ensures that the momentum distribution
is uniform in angle, and
for $r \gg l_\phi$ the energy is redistributed to ensure equilibrium at
the relevant (experimentally-measured) chemical potential of each terminal.
In the intermediate region, there is a broad cross-over from lead to reservoir.

% Outline..............................

In this work we first derive a general relation between
transmission cross section
(a concept we define using scattering in the half-plane)
and conductance for this open geometry, in Section~\ref{sec:rel}.
In Section~\ref{sec:pw} we show that partial-wave type states,
defined in the half-plane regions,
can take the place of transverse lead modes in the Landauer formula. 
In Section~\ref{sec:coupled} we
discuss the maximum conductance through an idealized, highly non-adiabatic QPC
(a hole in a thin hard wall) 
which is reached when a resonator is placed on one side of the QPC.
We find a universal result, namely a single conductance quantum,
{\em regardless} how small the hole is.
This illuminates the findings
of a recent experiment\cite{kati97} in such an open geometry.
In Section~\ref{sec:max} we discuss attempts to exceed this universal
quantum of conductance through a single channel.
A reciprocity relation for cross section is derived in Section~\ref{sec:atom},
and the possibility of breaking this
reciprocity, due to a non-thermal reservoir occupation, is described.
We discuss an application to matter-wave `conductance' through
a 3D QPC.
We conclude in Section~\ref{sec:conc}.

%ffffffffffffffffffffffffffffffffffffffffffffffffffffffffffffffffffffffff
%

%%%%%%%%%%%%%%%%%%%%%%%%%%%%%%%%%%%%%%%%%%%%%%%%%%%%%%%%%%%%%%%%%%%%%%%%%%%%%%%%
\section{Conductance in terms of cross section}
\label{sec:rel}

We consider scattering of a single-quasiparticle wavefunction
from the general 2-terminal system described in the
Introduction (see Fig.~\ref{fig:geom}a).
% definitions:
The Hamiltonian is ${\cal H} = -(\hbar^2/2m) \nabla^2 + V\ofr$,
for a quasiparticle mass $m$.
The elastic scattering potential $V\ofr$ completely defines the system.
We imagine a monochromatic unit plane wave
$\psi_{\text I} = e^{i\bk \cdot \br}$ incident
from the free-space left-hand region \cite{leftright}.
The wavevector is $\bk \equiv (k,\phi)$ in polar coordinates,
$\phi$ being the angle of incidence.
The free-space wavevector magnitude is taken as $k = \kf$
(corresponding to a total energy $E = \hbar^2 k^2 / 2m$
equal to the Fermi energy),
unless stated otherwise.

We are at liberty to choose our definition of the `unscattered' wave
$\psi_{\text 0}$.
We take it to be the wavefunction
which would result from reflection of the incident wave
off a wall uniform in the $y$ direction.
We can imagine creating such a wall by replacing the `system box' shown
in Fig.~\ref{fig:geom}a by the surrounding $y$-invariant
wall profile.
Note that $\psi_{\text 0}$ exists only on the left side.
In the left free-space region it is
\be
\label{eq:unscatt}
	\psi_{\text 0} \; = \;
	e^{i(k_x x + k_y y)} - e^{i(-k_x x + k_y y + \gamma_\bk)}
\ee
where the first term is $\psi_{\text I}$, and
the angle-dependent reflection phase $\gamma_\bk$ of the second term
depends on both $(k,\phi)$ and the wall profile
\cite{unscat}.
Upon introduction of our true system potential, the full wavefunction
becomes
\be
\label{eq:scat}
	\psi \; \equiv \; \psi_{\text 0} +  \psi_{\text R} + \psi_{\text T} ,
\ee
where the change in reflected wave $\psi_{\text R}$ exists only on the
left side, and the new transmitted wave $\psi_{\text T}$ exists only on the
right.
These scattered waves have the asymptotic ($r > L$ and $kr \gg 1$) forms
of 2D scattering theory \cite{scat},
\be
\label{eq:fdef}
	 \psi_{\text R} =
	f_{\text R}(\theta)\frac{e^{ikr}}{\sqrt{r}}, \hspace{.5in}
	 \psi_{\text T} =
	f_{\text T}(\theta')\frac{e^{ikr}}{\sqrt{r}}.
\ee
See Fig.~\ref{fig:geom}a for definitions of $\theta$ and $\theta'$.

The transmission cross section $\st(k,\phi)$
is the ratio of $\GT$, the transmitted particle flux (number per unit time), to
$j_{\text I}$, the incident particle flux per unit length 
normal to the incident beam:
\begin{equation}
\label{eq:sigmadef}
	\st(k,\phi) \; \equiv \; \frac{\GT}{j_{\text{I}}}.
\end{equation}
Physically, $\st(k,\phi)$ is the length required of an aperture
oriented normal 
to the incident beam in order
to transmit an equivalent flux of classical particles.
(Note that $\st(k,\phi)$ is proportional to the {\em injection distribution}
\cite{been} which can be measured in mesoscopic systems \cite{inj}).
It depends on the incident angle because $V\ofr$ has no radial symmetry.
$j_{\text{I}}$ is the magnitude of the incoming
probability flux density vector
${\bf j}  \equiv (\hbar/m) \text{Im} [ \psi_{\text I}^* \DEL \psi_{\text I} ]$,
which for a unit wave gives $j_{\text{I}} =  v$,
the particle speed.
The transmitted flux is defined as
\begin{equation}
\label{eq:GT1}
	\GT \; \equiv \;  \int \! dl \: \hat{\bf n}\cdot{\bf j} \; = \;
	\frac{\hbar}{m} \int \! dl \: \hat{\bf n}
	\cdot
	\text{Im} [ \psi_{\text T}^* \DEL \psi_{\text T} ] ,
\end{equation}
where the line integral encloses the entire transmitted wave,
and the (rightwards-pointing) surface normal is $\hat{\bf n}$.
Applying this and (\ref{eq:sigmadef}) to the asymptotic form gives
\be
\label{eq:sigmaf}
	\st(k,\phi) \; = \; \int_{-\pi/2}^{\pi/2} d\theta' \,
	|f_{\text T}(\theta')|^2 ,
\ee
familiar from scattering theory apart from the restriction
to the right half-plane.
There is a corresponding form
\be
\label{eq:sigmarf}
	\sigma_{\text R}(k,\phi) \; = \; \int_{-\pi/2}^{\pi/2} d\theta \,
	|f_{\text R}(\theta)|^2 ,
\ee
for the reflective cross section (removal from the unscattered wave
without being transmitted).

% ---------------begins actual calc of G:-----------------------------------
We will calculate the conductance by
assuming the chemical potential is slightly higher on the left side
than the right,
and as is usual\cite{been,datta}
consider only the left-to-right transport of the states in this
narrow energy range.
We take the left region to be a large ($ \gg l_\phi$) closed region
of area $A$
containing single-particle states,
and find their decay rate through the QPC into the
right side.
Semiclassically each single-particle state occupies a
phase-space volume $h^d$, where we have $d=2$.
Therefore the phase-space density in the 2DEG Fermi sea
is $2/h^2$ where the factor of 2 comes from the spin degeneracy.
We can project this density onto momentum space
in order to find the effective
number of plane-wave states impinging on the wall \cite{rectnote}:
this corresponds to a uniform density of states in $\bk$-space given by
\be
	\rho(k,\phi)\, k dk \, d\phi \; = \;
	\frac{A}{2 \pi^2} \, k dk \, d\phi .
\ee
Each state has an amplitude $A^{-1/2}$
due to the requirement of unity area normalisation in the left region,
so has incoming flux density $j_{\text I} = v/A$.
Substituting this into (\ref{eq:sigmadef})
gives the decay rate of a state $i$ as
\be
	\GT^{(i)} \; = \; \frac{v}{A} \st(k_i, \phi_i) .
\ee
We can now sum the decay rates of
all the left-hand states
in a given wavevector range $\kf$ to $\kf + \delta k$,
to get the current
\bea
\label{eq:cur}
	\delta I  =  e \sum_{i} \GT^{(i)} & = &
	\frac{ev}{A} \int_{-\pi/2}^{\pi/2} \!\! d\phi
	\int_{\kf}^{\kf + \delta k} \!\!\!\! kdk \,
	\rho(k,\phi) \, \st(k,\phi) \nonumber \\
	& = & \frac{ev \, \kf \delta k}{2\pi^2}
	\int_{-\pi/2}^{\pi/2} \! \! d\phi \, \st(\kf,\phi) ,
\eea
where the last step incorporated the linear-response assumption that
$\st$ is constant over the range $\delta k$.

When a potential difference $\delta V$ is applied across the QPC, the energy
range carrying current is $\delta E = e \, \delta V$,
which we can equate
with $\hbar v \, \delta k$ using the dispersion relation.
This can be used with (\ref{eq:cur}) to write the conductance
\begin{mathletters}
\label{eq:condgen}
\bea
\label{eq:condint}
	G \; \equiv \; \frac{\delta I}{\delta V} & = &
	\frac{2e^2}{h} \cdot \frac{1}{\lf}
	\int_{-\pi/2}^{\pi/2} \! \! d\phi \, \st(k_{\text F},\phi) \\
\label{eq:condavg}
	& = &
	\frac{2e^2}{h} \cdot \frac{k_{\text F}}{2}
	\langle \st \rangle_\phi ,
\eea
\end{mathletters}
where the particle wavelength is $\lf \equiv 2\pi/\kf$.
The latter form is written in terms of
the angle-averaged cross section at the Fermi energy.
The weighting of this average is uniform because of
the ergodic assumption that incoming states are uniformly distributed 
in angle.

Eq.(\ref{eq:condgen}) is a key result of this paper
(an independent derivation is given by Barnett~\cite{thesis}).
Like the Landauer formula, it directly connects
conductance and scattering.
In a scattering measurement from the left side, $\st$ appears to be
the QPC's inelastic
cross section (since the transmitted waves never return to this side).
In a current measurement the corresponding conductance is given by
(\ref{eq:condgen}).
Our derivation was for temperature $T=0$, but
it applies at a finite $T$ as long as $\st$ does
not change significantly over the energy range $k_{\text B} T$.
This can be seen by generalizing the above to include integration
over the Fermi distribution.

In the limit where a QPC is adiabatic, its conductance is known
to be quantized \cite{wees88,been,dittrich}:
$G = (2e^2/h)N$ where $N$ is the integer number of
open channels at the Fermi energy.
Looking at (\ref{eq:condint}),
this corresponds to
quantization of the angular integral of the cross section in units
of $\lf$.

%%%%%%%%%%%%%%%%%%%%%%%%%%%%%%%%%%%%%%%%%%%%%%%%%%%%%%%%%%%%%%%%%%%%%%%%%%
\section{Partial-wave channel modes for a 2-terminal system}
\label{sec:pw}

In free-space scattering
theory, partial waves
form a basis in which to decompose the asymptotic 
($r \rightarrow \infty$) form
of the full wavefunction $\psi$ into
incoming and outgoing states of definite angular momentum $l$.
In 2D the basis functions are the cylindrical solutions to the free-space
wave equation;
the $S$-matrix which takes incoming to outgoing waves
can then be written in this basis \cite{scat}.
Because there is only a single set of incoming channels
and a single set of outgoing channels, this is equivalent to a
scattering system (a `stub') connected
to a single `lead', with an infinite number of open channel modes.
This contrasts the
open two-terminal geometry we study,
where we need to account for two new related facts:
1) in the $r \rightarrow \infty$ limit the potential $V$ no longer
preserves angular-momentum, and
2) there are now distinct ways the particle can enter and exit the system,
via different leads.

We define a `half-plane partial-wave basis' as the subset of the cylindrical
free-space
%(Helmholtz)
solutions which go to zero on the entire $y$-axis.
This gives independent basis functions existing on either the left or right
side
of the $y$-axis. The basis is expressed in terms of
Hankel functions\cite{arfken} on either side
\bea
\label{eq:basis}
	\phi^{-L}_l (kr) &\; \equiv \;&
	H^{(2)}_l (kr) \sin [ l(\pit - \theta )]
	\nonumber  \\
	\phi^{+L}_l (kr) & \; \equiv \; &
	H^{(1)}_l (kr) \sin [ l(\pit - \theta )]
	\nonumber \\
	\phi^{-R}_l (kr) & \; \equiv \; &
	H^{(2)}_l (kr) \sin [ l(\pit - \theta' )]
	\nonumber \\
	\phi^{+R}_l (kr) & \; \equiv \; &
	H^{(1)}_l (kr) \sin [ l(\pit - \theta' )]
\eea
where on the left ($L$) side $\theta$ is the angle from the negative $x$-axis
and on the right ($R$) side
$\theta'$ is the angle from the positive $x$-axis
(see Fig.~\ref{fig:geom}a).
The channel index is $l = 1,2, \cdots \infty$, and
$+$($-$) refers to outgoing (incoming) travelling waves.
We note that the s-wave $l=0$ is excluded because of the $y$-axis barrier,
leaving the first channel as the p-wave $H_1(kr)\cos(\theta)$.
Assuming the width of the barrier is finite
and constant as $|y| \rightarrow \infty$ (see Fig. \ref{fig:geom}a), then
any wavefunction in the $r \rightarrow \infty$ limit
can be written as a sum of the above basis
functions.
The separability of this basis in $(r,\theta)$ is directly analogous to
the separability of conventional (constant-width) lead
basis states \cite{LB}
into a product of transverse modes and longitudinal travelling waves.

Our basis (\ref{eq:basis}) is chosen such that unit amplitude
coefficients carry equal fluxes
in all incoming and outgoing
channels, so flux conservation implies
the unitarity of the $S$-matrix when written in this basis.
As with a conventional transverse lead mode basis,
the familiar Landauer formula
\be
\label{eq:lb}
	G \; = \; \frac{2e^2}{h} \mbox{Tr}( t^\dag t) ,
\ee
holds\cite{stone,datta,thesis}.
The transmission matrix $t$
is defined by $q^+_l = \sum_m t_{lm} p^-_m$,
where the outgoing (incoming) amplitude coefficients
are $p^+_l$ ($p^-_l$) on the left and $q^+_l$ ($q^-_l$) on the right.
Note that is possible to `mix and match' different basis set types
(for instance
define a transmission matrix between transverse lead modes on the left side
and partial-wave modes on the right),
as long as equal-flux normalisation, and transverse
orthogonality, are preserved.

%%%%%%%%%%%%%%%%%%%%%%%%%%%%%%%%%%%%%%%%%%%%%%%%%%%%%%%%%%%%%%%%%%%%%%%%%%
\section{Point contact coupled to a resonator}
\label{sec:coupled}

\begin{figure}
\centerline{\epsfig{figure=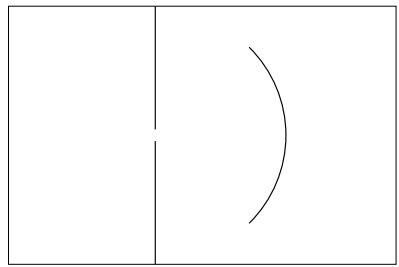,width=0.6\hsize}}
\vspace{0.1in}
\caption{
A tunneling-regime QPC combined with a nearby circular reflector, forming a
stable resonant cavity
open at the sides.
}
\label{fig:cavity}
\end{figure}

Fig. \ref{fig:cavity} illustrates a
QPC-plus-reflector system
whose conductance has been experimentally measured \cite{kati97}.
The circular arc reflector and the vertical wall
together form a cavity which can support
long-lived resonances;
the energy of these resonances can be swept by sweeping the reflector
gate voltage.
The classical condition\cite{kati97} for stability of the cavity modes is that
the arc center must lie at, or to the left of, the wall ($x{=}0$).
The cavity modes are coupled to the left terminal via the QPC, and
to the right terminal via leakage of the modes out through the
cavity top and bottom.
The system is interesting because it is `open' in the sense that
it has no Coulomb blockade \cite{been}, but `closed' in the sense that
the dwell time is much greater than the ballistic time (the resonances are
long-lived).
It has also been studied recently in our laboratory using microwave
measurements \cite{jesse}.

The actual potential in a mesoscopic experiment differs from
the illustration:
it has soft walls (on the scale $1/\kf$),
it may have deviations from the circle due to
lithographic error, and it has modulations of the background potential
due to elastic disorder \cite{kati97}.
However, we will not be interested in details of the
resonator on the right-hand side.
Rather, we will adopt the view of a 2D scattering-theorist
`looking' from the left-hand side.
In this section we discuss the maximum
conductance of this system, when the `bare' QPC (\ie without the reflector)
is in the tunneling regime (conductance $ \ll 2e^2/h$).

%fffffffffffffffffffffffffffffffffffffffffffffffffffffffffffffffffffffffffffff

We use an idealized slit QPC model (see Fig~\ref{fig:geom}b)
in which the potential $V$ is zero
everywhere 
except along a hard, thin wall where it is taken as infinite.
The QPC is a gap in the wall of size $2a$.
This model is highly non-adiabatic
(see Ref.~17 for a review
of its transmission properties).
The hard wall
simplifies the treatment of the left-hand side scattering problem,
and we do not believe it alters our
basic conclusion.
We consider the `unscattered'
wave to be the incident plus reflected wave Eq.(\ref{eq:unscatt})
when the QPC is closed ($a=0$).
This we expand in Bessel functions,
\bea
\label{eq:expand}
\psi_{\text 0}\ofr & =  & e^{i(k_{x} x + k_{y} y)} - e^{i(-k_{x} x + k_{y} y)} 
\nonumber \\
& = & -4 i J_{1}(kr)\cos(\theta)\cos(\phi) + 
\mbox{\rm higher order terms}.
\eea
The first term in the expansion is the incoming plus outgoing p-wave,
which in the tunneling limit will dominate
in our consideration of the absorption \cite{thesis}.

Now we open the slit, and replace
$2 J_{1} (kr)$ in the above by
$H_1^{(2)}(kr) + e^{2i\delta}\, H_{1}^{(1)}(kr)$,
where $\delta$
% (which depends on the whole scattering geometry)
follows the usual definition of partial-wave phase shift \cite{scat}.
The closed slit corresponds to $\delta = 0$.
An open slit leading into a closed resonator
(imagine extending the arc in Fig.~\ref{fig:cavity} to seal off the
entire right side), in the case of infinite dephasing length,
corresponds to $\delta = $ real, and would appear from the left side
as an elastic dipole scatterer.
An open slit with an open resonator
corresponds to complex $\delta$ with positive imaginary part,
and would appear as a general inelastic dipole scatterer.
Therefore transmission though the QPC appears,
to an observer on the left side, to be {\em absorption} of incident waves.
$\st$ is interpreted as an `inelastic' cross section (since exiting
the right-hand terminal is equivalent to leaving in a new channel),
and $\sigma_{\text R}$ as an `elastic' one.
$\sigma_{\text T}(k,\phi)$ can be
found from integrating the net incoming flux [as in Eq.(\ref{eq:GT1})]
of the total wavefunction on the {\em left} side.
Substitution into (\ref{eq:sigmadef}) then gives
$\sigma_{\text T}(k,\phi) = \frac{4}{k} (1 - |e^{2i\delta}|^2) \cos^2(\phi)$ .
For $\delta \rightarrow i \infty$ the maximal cross section is reached,
\be
\label{eq:maxxsec}
	\sigma_{\text{T,max}}(k,\phi) \; = \;
	\frac{4}{k} \cos^2(\phi) .
\ee
This corresponds to an effective classical `area' (size) $\aeff = \lf/2$.
This is analogous to the fact\cite{jackson} that in 3D
the effective area of an arbitrarily-small
electromagnetic dipole aerial can be of
order $\lambda^2$.
To an observer on the left side who was able to `see' the
electron waves living in the energy range $e \, \delta V$
responsible for conductance,
the QPC would stand out as a `black dot' of size $\sim \lf$
against the surrounding uniform `grey' thermal luminosity reflected
in the vertical wall mirror.

The associated maximum conductance is found easily
using (\ref{eq:maxxsec}) and (\ref{eq:condgen}) to be
\be
\label{eq:maxg}
	G_{\text{max}} \; = \; 	\frac{2e^2}{\hbar},
\ee
the universal quantum of conductance (for 2 spin channels),
independent of the size of the QPC hole, even for an
{\em arbitrarily small} hole ($ka \rightarrow 0$).
This universal resonant-tunelling maximum conductance
was first found numerically
\cite{xue,kalm,been}; however our system
differs from those of Xue \etal \cite{xue} and
Kalmeyer \etal \cite{kalm} because the resonance does not
involve transmission though an {\em isolated}
(zero-dimensional) quantum dot.
The dramatic increase over the conductance of the bare QPC
(which vanishes as $(ka)^4$, see Ref.~\cite{thesis})
runs counter to the naive classical expectation, namely
that the reflector would {\em decrease} the left-to-right flow of electrons
because it sends back into the QPC particles which would otherwise
exit to the right.

How do we know that it is possible to build a resonant geometry
which corresponds to $\delta \rightarrow i \infty$?
The reflector can be described by $r$, the amplitude with which it returns
an outgoing p-wave back to the QPC as an incoming p-wave.
If $|r|^2 = 1 - |t_{11}|^2$, where the p-wave transmission of the QPC
is $t_{11}$ as defined in Section~\ref{sec:pw}, then the p-wave channel
becomes a 1D Fabry-Perot resonator with mirrors of matched
reflectivity.
Sweeping the round-trip phase then produces peaks of complete transmission
(corresponding to complete p-wave absorption on the left side).
The ratio of peak separation to peak width is the quality factor
$Q \sim 1/|t_{11}|^2$.
Such peaks, with heights much greater than the bare tunneling QPC conductance,
were observed in the experiments
of Katine {\it et al.}\cite{kati97}.
However, Eq.(\ref{eq:maxg}) has not yet been tested quantitatively because
of the difficulty of matching the Fabry-Perot reflectivities in a real
2DEG experiment.
Note that the maximum conductance (\ref{eq:maxg}) also
follows immediately from
the Landauer formula when we realize that there can be complete transmission
of the incoming $l{=}1$ channel state (from Section~\ref{sec:pw}).

An interesting possibility arises when we realize~\cite{thesis}
that higher $l$ channels are still {\em slightly} transmitted by the bare QPC,
when $ka \ll 1$, even though they are increasingly evanescent.
If the resonator has a high enough reflectivity for these modes, then
additional Fabry-Perot conductance peaks will be produced \cite{xue,bryant}.
The peaks may be extremely narrow, but can carry a full quantum of conductance
because they can transmit another incoming $l$ channel.
By careful arrangement of the cavity, one or more of these
peaks could be brought into conjunction with an already-existing
$l{=}1$ peak at the Fermi energy.
(For instance, the $l{=}1$ and $l{=}2$ resonances are in different
symmetry classes in Fig.~\ref{fig:cavity}
so there can be an exact level crossing).
Therefore, we have the surprising result that, in theory, a
conductance of $(2e^2/h) n$ can pass through
an arbitrarily small QPC hole if $n$ resonances (from $n$ different channels)
coincide at the Fermi energy.
However, due to their extremely small width, such large conductance
peaks are unlikely to be observable in a real mesoscopic tunneling QPC
due to finite dephasing length and finite-temperature smearing \cite{been}.

Finally, we should not overlook the fact that our expressions for
partial cross sections are a factor of 4 greater
than those conventionally arising in 2D scattering theory
from a radial potential \cite{scat},
because we are measuring cross section on the reflective boundary of a
semi-infinite half plane.
For instance, the maximum inelastic partial cross section for a single
channel in free space \cite{scat} is $\sigma_r = 1/k$, compared to our
maximum `inelastic'
cross section per channel Eq.(\ref{eq:maxxsec}).
Similarly, the maximum elastic result in free space is $\sigma_e = 4/k$,
compared to our maximum (normal-incidence) `elastic' cross section
per channel $\sigma_{\text{R,max}} = 16/k$.
This latter case occurs when $\delta = (\mbox{integer} + \half) \pi$.

%fffffffffffffffffffffffffffffffffffffffffffffffffffffffffffffffffffffffffffff
\begin{figure}
\centerline{\epsfig{figure=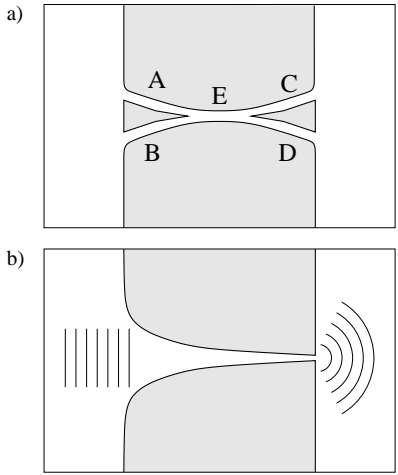,width=0.6\hsize}}
\vspace{0.1in}
\caption{
{\bf a)}
An attempt to increase conductance through a single channel by multiple
connections feeding from the reservoirs.
All channels are single-mode and sufficiently long that the evanescent
tunneling of higher modes is negligible.
{\bf b)}
An illustrative hard-walled exponential horn system which has differing
acceptance angles on each side:
very narrow on the left, and very wide on the right.
Such a mesoscopic 2DEG system would exhibit symmetric
conductance, however, in an atom beam context the conductance
can become unsymmetric.
}
\label{fig:multi}
\end{figure}

%%%%%%%%%%%%%%%%%%%%%%%%%%%%%%%%%%%%%%%%%%%%%%%%%%%%%%%%%%%%%%%%%%%%%%%%%%%%%%%%
\section{What is the maximum conductance of a single quantum channel?}
\label{sec:max}

The surprising theoretical results of the previous section might lead one to
question the conductance limit $2e^2/h$ for a single quantum channel
(by which we mean a single transverse mode for which the longitudinal
degree of freedom is a 1D Fermi gas; this includes both conventional
and partial-wave basis sets).
For this gedanken-experiment we will consider conventional
electron waveguides which are single-mode
and long enough that evanescent waves are negligible, but which are also
$\ll l_\phi$.
We try to encourage more current to pass down a single-mode channel (E)
by connecting it to a reservoir via multiple routes (A,B,C,D), as shown
in Fig~\ref{fig:multi}a, where two routes are used on each side.
It is possible to match the junctions so that a wave entering
down A,B,C, or D has no reflection back along the same lead.
In this case we might guess that the hypothetical left-side observer
(from the previous section) would see the
single-mode entrances to guides A and B as two `black dots', giving
twice the effective absorption cross section, and therefore infer a conductance
of twice $2e^2/h$.
We might also justify this by saying that waves travelling down A and B
will meet and continue down E, and
since they have no particular phase relation,
their currents
will add to give a doubled current through E, as would be necessary.

However there is a fundamental flaw in the above reasoning.
The ABE junction can be
designed so that if waves come down A and B in phase, they will be
adiabatically transformed into the lowest transverse mode of E,
so will propagate through to the right side without reflection, carrying
a current of twice that of a usual single-mode guide.
However, if A and B are $\pi$ out of phase, the same adiabatic transformation
must result in the second transverse mode, which is evanescent.
So this latter wave will reflect perfectly back out of the left side,
and carry no current.
Plane waves are impinging from the left reservoir uniformly over all angles,
and because of the $> \lambda$ separation of the entrances, an average
over angles gives an average over relative phase in A and B.
Thus we are left with no increase above the single channel conductance.
This property of the ABE junction is not merely practical;
rather, it is easy to show that its $3\times3$ $S$-matrix
cannot be unitary
if a junction is to couple both A$\rightarrow$E and B$\rightarrow$E
with unity transmissions.
Such an appealing junction is therefore
ruled out on the grounds of flux conservation.
A consequence is that the entrances to A and B can at most
appear `half black' to the observer, due to waves which enter A then
exit B and vice versa.
%Hence the effective absorption cross section is no more than that
%of a single channel.
%
%Argument of seeing two black dots when look at 2 slits coupled together?
%Cannot since if close enough to resolve dots,
%each will only look half-dark (stuff comes shining out which was sent down
%the other arm).
%If far away, cannot resolve dots and get an incident-angle oscillation
%from 2 black dots to 0 black dot intensity.

This suggests another way to try and defeat the conductance limit:
direct the incoming plane waves in a narrow enough angular distribution
so that waves {\em always} come down A and B in phase, and this
will double the conductance.
(This is similar to experiments \cite{series} where the series resistance
of two QPCs was found to be less than the sum of the individual QPC resistances,
because collimation at the exit of the first QPC illuminated the
second with a narrow beam, increasing its conductance).
However, this beam is no longer a {\em thermal} occupation of incoming
states.
This illustrates the inextricable link between thermal Fermi occupation
of reservoir states and the universal quantum of conductance.
At $T{=}0$, thermal occupation at a given chemical potential difference
implies that {\em all} quantum states
lying in the appropriate energy range
are filled in the left reservoir and empty in the right.
Semiclassically, this corresponds to a uniform distribution in phase space,
or when projected into momentum states, uniform in angle, as exemplified
by Eq.(\ref{eq:condgen}).
The semiclassical viewpoint allows one to see that since
transformations in phase space cannot change the phase space density
(Liouville's theorem), neither can the universal  
conductance per quantum channel be changed.
This reminds us that unitarity in quantum mechanics
is analogous to Liouville's theorem in classical mechanics.

%%%%%%%%%%%%%%%%%%%%%%%%%%%%%%%%%%%%%%%%%%%%%%%%%%%%%%%%%%%%%%%%%%%%%%%%%%%%%%%%
\section{Reciprocity and `conductance' of atom waves}
\label{sec:atom}

We can ask if the conductance (\ref{eq:condgen}) computed using
transmission of left-side reservoir plane wave states through the QPC is equal
to that using right-side reservoir states.
Since the two directions correspond to opposite signs of $\delta V$,
then in order to have linear response (well-defined
constant $G$ around $\delta V = 0$)
we would hope that they are equal.
That the angular average of transmission cross section is equal
from the left and right sides is not immediately apparent in a general
asymmetric system.
For instance, consider Fig.~\ref{fig:multi}b which has a small acceptance
angle from the left but a large from the right, therefore
very different forms of the
transmission cross sections $\stlr (k,\phi)$ and $\strl (k,\phi)$.

If we assume {\em classical}
motion then we can imagine a map from a Poincar\'{e}
Section (PS) $(y,p_y)$ at a vertical slice at $x = -x_0$ to another PS
$(y',p_y')$ at $x = +x_0$.
At each PS we consider only rightwards-moving ($p_x > 0$) particles, and
take $x_0 > L$.
A certain area of phase space $(y,p_y)$
is transmitted and is mapped
to an {\em equal area} \cite{classmech} in phase space $(y',p_y')$.
Time-reversal invariance holds since we consider magnetic field $B{=}0$,
so we can negate the momenta (now considering $p_x < 0$) and find that
the {\em same} phase space area is transmitted right-to-left.
When it is realised that the angle-averaged cross section is proportional
to the transmitted phase space area on a PS, then the symmetry
of the angle-averaged {\em classical} cross sections follows.

The same symmetry is not obvious for quantum cross sections, but
it also holds true.
Comparing (\ref{eq:condint}) with (\ref{eq:lb}) gives
\be
\label{eq:cool}
	\int_{-\pi/2}^{\pi/2} \! \! d\phi \, \stlr(k_{\text F},\phi)
	\; = \;
	\lf \mbox{Tr}( t^\dag t) ,
\ee
where $t$ is measured from left to right states.
It is instructive to derive this directly \cite{thesis}.
This relation ties together the cross section
and Landauer views of conductance.
Time-reversal invariance
and flux conservation together imply \cite{magnetic}
that $\mbox{Tr}( t^\dag t)$
is unchanged by swapping the labelling of the leads \cite{been,stone}, thus
we immediately have from (\ref{eq:cool}) the reciprocity of
angle-integrated quantum cross section
\be
\label{eq:recip}
	\int_{-\pi/2}^{\pi/2} \! \! d\phi \, \stlr(k,\phi)
	\; = \;
	\int_{-\pi/2}^{\pi/2} \! \! d\phi \, \strl(k,\phi) .
\ee
So in Fig.~\ref{fig:multi}b is it now
clear that the ratio of acceptance angles
must be balanced by the ratio of effective areas.

We now discuss a case in which non-thermal occupation
of incoming states is possible: the
rapidly developing field of coherent matter-wave optics, in which
potentials are defined by microfabricated structures
\cite{chips,joseph,me}.
There is a recent proposal \cite{joseph} for observation of
quantization of atomic flux
through a micron-sized 3D QPC defined by the Zeeman effect potential of
a magnetic field.
The device is illuminated by a beam of atoms passing through a vacuum,
whose angular distribution is an experimental parameter
(for instance, a collimated oven source or a dropped cloud of cold atoms
\cite{dropped}).
The atomic flux transmitted (per unit $k$, at wavevector $k$) will be
$F(k) = \gat(k) \, J_{\text 0}(k)$
where $J_{\text 0}(k)$ is the flux incident per unit wall area,
and we define the atomic `conductance' by
\be
\label{eq:atomcond}
	\gat(k) \; \equiv \; \int d\Omega \, w(k,\Omega) \, \st(k,\Omega) .
\ee
As before, the quantum transmission cross section is $\st(k,\Omega)$,
but now there is a {\em weighting function} $w(k,\Omega)$ which defines
the angular distribution of the incident beam \cite{brightnote}.
The weight has the normalization
$\int d\Omega \, w(k,\Omega) \, \cos(\theta) = 1$.
[All integrals over solid angle $\Omega \equiv (\theta,\phi)$ are over a
range of $2\pi$ appropriate for the half-sphere].
Following the analogy of Thywissen\cite{joseph},
$F(k)$ plays the role of current, $J_{\text 0}(k)$ that of bias voltage.
However, the name `conductance' does not imply any definite chemical
potential difference as in the 2DEG case. 
For classical particles, the `conductance' of an aperture of
area $\aaeff$ in a thin wall is simply $\gat(k) = \aaeff$,
regardless of the incident angular distribution.
Thus $\gat(k)$ gives the effective area $\aaeff$
of a QPC, in an analogous fashion
to $\aeff$ in 2D.

For an integer number of quantum channels, the
2D quantization of $\aeff$ in units of $\lambda/2$ 
becomes in 3D
the quantization\cite{joseph,been,sharvin}
of $\aaeff$ in units of $\lambda^2/\pi$,
a result well known from
work on 3D metallic point contacts\cite{krans}.
As stated by Thywissen\cite{joseph}, this accurate flux quantization
requires the incident beam width to be much larger than the QPC acceptance
angle.

Eq.(\ref{eq:atomcond}) is the matter-wave equivalent of
Eq.(\ref{eq:condint}), with the
important difference that it has a general weight function.
Possible non-uniformity of this weight function leads to a key result:
that {\em asymmetry} of the conductance is possible given
identical illumination on either side, even though the (center of mass)
motion is time-reversal invariant.
For example, if the incident flux used to illuminate
the horn QPC of Fig.~\ref{fig:multi}b
is narrow in angular spread,
then the left-to-right conductance will be much larger
than the right-to-left conductance.
This contrasts with the 2DEG case where the conductance is
always symmetric.

Finally, it is interesting to note that for the non-thermal
incident (reservoir) distributions
discussed above, the 
Landauer formula takes the modified form
\be
\label{eq:modlb}
	G \; \propto \; \mbox{Tr} (t^\dag t \rho)
\ee
where $\rho$ is the density matrix of the incident
beam.

%%%%%%%%%%%%%%%%%%%%%%%%%%%%%%%%%%%%%%%%%%%%%%%%%%%%%%%%%%%%%%%%%%%%%%%%%%%%%%%%
\section{Conclusions}
\label{sec:conc}

Quantum scattering theory in the 2D half-plane
can provide an alternative description of the mesoscopic
conductance of non-interacting particles.
It is especially useful in `open'
systems (\eg those with nearby scatterers in the reservoir regions)
where the usual transverse-channel approach is inappropriate.
We have considered elastic potentials in
zero magnetic field, in linear response in the low temperature limit.
Conductance is proportional to the transmission 
cross section integrated over all incident angles, Eq.(\ref{eq:condgen}).
We also define a half-plane partial-wave basis
applicable with the usual Landauer formula, and relate this to our transmission
cross section result.
A difference between this and previous work is the ability to treat
a direct `leadless' connection to the reservoir.

Using the example of a slit QPC combined with an open
cavity structure,
we show that an arbitrarily small QPC can carry up to a single
quantum of conductance via resonant tunnelling (equal to the limit in
the closed-dot resonant tunnelling case).
This requires a resonance at the Fermi energy.
If $n$ coincident resonances occur for different incoming channels,
then $n$ conductance quanta can in theory be achieved through this same
tunneling QPC, a result which we believe has not been noted until now.

We emphasize that conductance is proportional
to phase-space density of the reservoir states.
Therefore
the universal quantum of conductance $e^2/h$ per spin in Fermi gas systems
is a direct result of the uniform phase-space density
(angular distribution) in a thermal occupation of the Fermi sea.
This insight is supported by discussion of attempts to exceed
this universal value.
When the reservoir occupation differs from thermal, the
conductance formula requires generalization: an angle-dependent weight
is included in the cross section integral (\ref{eq:atomcond});
equivalently for 2DEG systems the Landauer
formula requires inclusion of the incoming ensemble (\ref{eq:modlb}).
This result, and our approach in general,
is relevant to the emerging field of matter-wave conductance by microfabricated
structures (for instance, a quantum point contact in 3D),
under general illumination by atom waves.
We hope this work provides new tools for the study of coherent 
electron and matter-wave systems.

%kkkkkkkkkkkkkkkkkkkkkkkkkkkkkkkkkkkkkkkkkkkkkkkkkkkkkkkkkkkkkkkkkkkkkkkkkk
\acknowledgments
We thank Adam Lupu-Sax
for his early contributions to this work,
also J. Thywissen, C. Marcus and D. Fisher for stimulating discussions.
We also thank our referees for alerting us to additional references.
This work was supported by the National Science Foundation (USA)
on grant no.\ CHE-9610501,
ITAMP (at the Harvard-Smithsonian Center for Astrophysics and the Harvard
Physics Department), and
the Netherlands Organization for Scientific Research (NWO).

%rrrrrrrrrrrrrrrrrrrrrrrrrrrrrrrrrrrrrrrrrrrrrrrrrrrrrrrrrrrrrrrrrrrrrrrrrrrrrrr


\begin{references}

\bibitem{been}
%H. van Houten and C.W.J. Beenakker, Physics Today, July 1996;
For a review see C.W.J. Beenakker and H. van Houten, 
Solid State Physics {\bf 44}, 1 (1991).

\bibitem{dittrich}
T. Dittrich, P. H\"{a}nggi, G.-L. Ingold, B. Kramer, G. Sch\"{o}n, W. Zwerger,
{\it Quantum Transport and Dissipation}, (Wiley-VCH, Weinheim, 1998).

\bibitem{LB}
R. Landauer, IBM J. Res. Dev. {\bf 1}, 233 (1957); 
{\it ibid.} Z. Phys. B {\bf 68}, 217 (1987); M. B\"{u}ttiker,
Phys. Rev. Lett. {\bf 57}, 1761 (1986).

\bibitem{datta}
S. Datta, {\it Electronic Transport in 
Mesoscopic Systems}, (Cambridge University Press, NY, 1995).

\bibitem{wees88}
B.J. van Wees {\it et al.}, Phys. Rev. Lett. {\bf 60}, 848 (1988).

\bibitem{nonadiab}
By non-adiabatic, we mean that even at a QPC's narrowest region
the transverse profile is changing rapidly.
Clearly every QPC becomes `non-adiabatic' at the coupling to
infinite-width reservoirs:
this type of non-adiabaticity we do not include because it
does not cause significant impedance mismatch, as explained by Yacoby
and Imry\cite{yacoby}.

\bibitem{yacoby}
A. Yacoby and Y. Imry, Phys. Rev. B {\bf 41}, 5341 (1990).

\bibitem{kati97}
J. A. Katine {\em et al.}, Phys. Rev. Lett. {\bf 79}, 4806 (1997).

\bibitem{topinka}
M. A. Topinka, {\it et al.} Science {\bf 289}, 2323 (2000).

\bibitem{geim94}
A.K. Geim {\it et al.}, Phys. Rev. B {\bf 49}, 2265 (1994). 

\bibitem{szafer}
Although the quasi-1D approach can be retained by modelling very wide
leads attached
to such systems, following A. Szafer and A. D. Stone, Phys. Rev. Lett.
{\bf 62}, 300 (1989),
this has both numerical and conceptual limitations.

\bibitem{leftright}
Of course, throughout this paper
we could imagine the incident wave on the right-hand side,
and the same conductance would result (since we are in linear
response); see Section~\ref{sec:atom}.

\bibitem{unscat}
We could equally well
imagine that the QPC can be `closed off'
(no transmission) by varying a parameter
(this is often true experimentally), and define $\psi_{\text 0}$
as the full wavefunction in this closed-off state.
Thus $\psi_{\text 0}$ would be the sum of an incident plane wave
and a more complicated outgoing wave.
This alternative definition may be better in
systems where the wall has disorder, or where there is more
complicated structure
on the left-hand side than shown in Fig.~\ref{fig:geom}a.
The two definitions are equivalent as far as Section~\ref{sec:coupled} is 
concerned.

\bibitem{scat}
%R.G. Newton, {\it Scattering of waves and particles}, 
%(Springer, New York, 1982)
L. D. Landau and E. M. Lifshitz, {\it Quantum Mechanics}, (Elsevier, 1998);
J. J. Sakurai, {\it Modern Quantum Mechanics, Revised Edition},
(Addison-Wesley, 1994), Chapter 7.
Presentations of two-dimensional scattering theory are rare;
see
I. R. Lapidus, Am. J. Phys. {\bf 50}, 45 (1982);
S. K. Adhikari, Am. J. Phys. {\bf 54}, 362 (1986);
Y. S. Chan, Ph.~D. thesis, Harvard University, 1997.

\bibitem{inj}
K. L. Shepard, M. L. Roukes, B. P. Van der Gaag, Phys. Rev. Lett. {\bf 68},
2660 (1992).

\bibitem{rectnote}
This argument can also be verified in the more specific case
of the left region being a
rectangular Dirichlet box, in which case the exact eigenfunctions
are known and can be written explicitly in terms of a sum of
$\psi_{\text 0}$ for incidences $\phi$ and $-\phi$.
However the phase-space presentation is more general, and applies
to the real situation where the left region is chaotic (diffusive
elastic scattering).

\bibitem{thesis}
A. H. Barnett, Ph.D. thesis, Harvard University, 2000.

\bibitem{arfken}
G. Arfken, {\it Mathematical Methods for Physicists,
2nd Edition}, (Academic Press, 1985).

\bibitem{fisherlee}
D. S. Fisher and P. A. Lee, Phys. Rev. B {\bf 23}, 6851 (1981).

\bibitem{stone}
A. D. Stone and A. Szafer, IBM J. Res. Develop. 
{\bf 32}, 317 (1988).


%\bibitem{abra65}
%{\it Handbook of mathematical functions}, Ed. by M. Abramowitz and I.S. Stegun, 
%(Dover, New York, 1965), Ch. 20.

\bibitem{jesse}
J. S. Hersch, M. R. Haggerty, and E. J. Heller,
Phys. Rev. Lett. {\bf 83}, 5342 (1999);
Phys. Rev. E. {\bf 62}, 4873 (2000).

\bibitem{jackson}
J. D. Jackson, {\em Classical Electrodynamics}, (Wiley, N.Y., 1975).

\bibitem{xue}
W. Xue and P. A. Lee, Phys. Rev. B {\bf 38}, 3913 (1988).

\bibitem{kalm}
V. Kalmeyer and R. B. Laughlin, Phys. Rev. B {\bf 35}, 9805 (1987).

\bibitem{bryant}
G. W. Bryant, Phys. Rev. B {\bf 39}, 3145 (1989).

\bibitem{series}
D. A Wharam, M. Pepper, H. Ahmed, J. E. F. Frost, D. G. Hasko, D. C. Peacock,
D. A. Richie, and G. A. C. Jones, J. Phys. C {\bf 21}, L887 (1998);
also see the review\cite{been} and references within.

\bibitem{classmech}
M. C. Gutzwiller, {\it Chaos in Classical and Quantum Mechanics},
(Springer-Verlag, N.Y., 1990), Ch. 7.

\bibitem{magnetic}
With $B{\neq}0$, the conductance is {\em still}
symmetric under swapping the leads.
This results from the 2-terminal special case of
unitarity sum rules\cite{datta}, namely that
the rows and columns of the matrix of absolute-value-squared $S$-matrix
elements
must all sum to 1.
Thus the reciprocity derived here is preserved for $B{\ne}0$.
How the classical argument from the previous paragraph generalizes for
$B{\neq}0$ is not known by the authors.

\bibitem{chips}
R. Folman, P. Kr\"{u}ger, D. Cassettari, B. Hessmo, T. Maier,
and J. Schmiedmayer, Phys. Rev. Lett. {\bf 84}, 4749 (2000).

\bibitem{joseph}
J. H. Thywissen, R. M. Westervelt, and M. Prentiss,
Phys. Rev. Lett.
{\bf 83}, 3762 (1999).

\bibitem{me}
A. H. Barnett, S. P. Smith, M. Olshanii, K. S. Johnson, A. W. Adams, and
M. Prentiss, Phys. Rev. A. {\bf 61}, 023608 (2000).

\bibitem{dropped}
M. Key {\it et al.}, Phys. Rev. Lett. {\bf 84}, 1371 (2000).

\bibitem{brightnote}
Because we wish to consider general illumination and general $st(k,\Omega)$,
our definition of `conductance' coincides with that of Thywissen~\cite{joseph}
only in the case of isotropic illumination $w(k,\Omega) = 1/\pi$.
The beam {\em brightness} per unit $k$ range, that is, its
phase-space density, is assumed uniform in position space,
and is proportional to $J_{\text 0}(k) \, w(k,\Omega)$.
This is also proportional to
$a(\bk)$ defined by Thywissen.

\bibitem{sharvin}
Yu. V. Sharvin, J. Exptl. Theoret. Phys. (U.S.S.R.) {\bf 48}, 984-985, (1965)
[trans. in Sov. Phys. JETP {\bf 21}, 655 (1965).]

\bibitem{krans}
J.M. Krans, J. M. van Ruitenbeek, V. V. Flsun,
I. K. Yanson, and L. J. de Jongh, Nature {\bf 375}, 767 (1995);
P. Garc\'{\i}a-Mochales, P. A. Serena, N. Garc\'{\i}a, and
J. L. Costa-Kr\"{a}mer, Phys. Rev. B {\bf 53}, 10268 (1996).


\end{references}
\end{document}